# Single metallic nanoparticle imaging for protein detection in cells

L. Cognet, C. Tardin, D. Boyer, D. Choquet[*], P. Tamarat and B. Lounis

*Centre de Physique Moléculaire Optique et Hertzienne - CNRS UMR 5798 et Université Bordeaux 1, 351 Cours de la Libération, 33405 Talence, France*

[*]*Laboratoire de Physiologie Cellulaire de la Synapse - CNRS UMR 5091 et Université Bordeaux 2, Institut François Magendie, 1 rue Camille Saint-Saëns 33077 Bordeaux, France*

Classification: Biological Science (Biophysics)

**Corresponding author:**

Prof. B. Lounis
Centre de Physique Moléculaire Optique et Hertzienne
CNRS UMR 5798 et Université Bordeaux 1
351 Cours de la Libération
33405 Talence, France
b.lounis@cpmoh.u-bordeaux1.fr

**Manuscript information :**

20 text pages (including title page, abstract, text, acknowledgements, references, figure legends),
4 pages of figures,
Supporting information
1 page
1 movie file.

**Word and character counts :** 97 words in the abstract, 31966 characters in paper

**ABSTRACT**

We performed the first visualization of membrane proteins labeled with 10nm gold nanoparticles in cells, using an all optical method based on photothermal interference contrast (PIC). The high sensitivity of the method and the stability of the signals allows 3-dimension imaging of individual nanoparticles without the drawbacks of photobleaching and blinking inherent to fluorescent markers. A simple analytical model is derived in order to account for the measurements of the signal amplitude and the spatial resolution. The PIC method provides an efficient, reproducible and promising way to visualize low amounts of proteins in cells by optical means.

An important development in modern optical microscopy is the direct observation of single molecules(1), in media as varied and complex as molten polymers(2) or living cells(3-7). Selecting a single molecule at a time allows the elimination of all the implicit averages of conventional optical observations, gaining access to heterogeneity, dynamical fluctuations(8, 9), diffusion(10), reorientation(11), colocalization(12), and conformational changes(13) at the molecular level. Single molecule methods have changed the way we think and carry out experiments on complex molecular systems. Yet, the optical detection of a single molecule remains difficult, because the spatial resolution is limited by the wavelength of light, so that the signal must be extracted from the background arising from billions of other molecules in the focal spot of the microscope. An elegant way to find the needle in the haystack is to attach a "label" to the molecule of interest. The ideal label must fulfill contradictory requirements. It should generate an intense optical signal, but at the same time be as small as possible, not to perturb the observed molecule too severely.

Several kinds of optical labels have been developed and utilized in recent years. The most commonly used labels are fluorophores, usually organic dyes that can be chemically grafted to the molecule under study, but also autofluorescent proteins genetically fused to the protein of interest. Because of the wavelength change it involves (Stocke's- shift), fluorescence can be separated from background. Usual dye fluorophores are so small that they hardly hamper the diffusion of their host proteins or lipids, nor do they affect their functions or interactions with partners, at least if they are not attached to sensitive areas. The main drawbacks of organic fluorophores are the photobleaching and the blinking, i.e. the processes by which photochemical reactions transform the excited fluorophore into a non-fluorescent product. A fluorophore's lifetime is severely limited by the heavy laser irradiations used in single molecule detection. Photobleaching can be reduced by removing free oxygen, but is never eliminated completely, and it is the most serious obstacle to many experiments such as in living cells due to the toxicity of the oxygen scavengers.

Nanocrystals of II-VI semiconductors (such as CdSe) have recently been used as fluorescent markers (14-17). Nanocrystals of different sizes can be excited by the same laser, and still be distinguished by their size-dependent luminescence spectrum. They must be passivated, protected and functionalized by surface layers, because of their high chemical reactivity, and in order to prevent luminescence quenching by acceptor surface states. Therefore, a passivated nanocrystal is a rather bulky label of several nanometers in diameter. Although they resist bleaching longer than organic dyes, their luminescence is subject to blinking(18), and they also bleach eventually.

Metal particles are commonly used for single-particle or single-molecule tracking(19) and in immunocytochemistry(20, 21). They are often colloids, with diameters ranging between a micron and a few nanometers, or chemically synthesized clusters(22, 23) with well-defined numbers of atoms and chemical structures. Metal particles are very appealing labels because they do not photobleach, and do not optically saturate at reasonable exciting intensities. Single large metal particles can be imaged in optical microscopy by means of various methods, such as imaging at the plasmon frequency with dark field illumination(24), with differential interference contrast (DIC) and video enhancement(25), or with total internal reflection(26). All these methods are based on the intense Rayleigh scattering of sub-micron particles, down to a diameter of about 40 nm. However, they do not apply to very small metal particles, because Rayleigh scattering decreases like the sixth power of diameter. Moreover, the scattering signal must be discriminated from a strong background, specially if the particles are to be detected in cells or tissues. Recently, we developed a photothermal interference contrast (PIC) technique to optically detect absorbing nanoparticles(27). The principle of the PIC method relies on the strong optical absorption of a small metal particle at their plasma resonance which gives rise to a photothermal effect, i.e. a change in temperature around the particle when it is illuminated by a laser light (heating laser). This temperature change leads to a variation of the local index of refraction which can be optically detected when combining high-frequency modulation and polarization interference contrast using a second laser beam. We could previously image very small gold colloids down to 5 nm or 2.5 nm in diameter embedded in thin polymer films, with a signal-to-noise ratio better than 10 or of the order of 2 respectively. We also showed that, in addition to its intrinsic sensitivity, the photothermal image is remarkably insensitive to scattering background, even when arising from such strong scatterers as 300 nm latex beads.

In the present work, we show that it is possible to image gold colloids in thick samples (e.g. cells) with a modified PIC setup. We demonstrate imaging of receptor proteins stained with individual 10 nanometer size gold particles in the plasma membrane of COS7 cells. The resolution in 3D of the PIC method is also derived and confronted to measurements.

## METHODS

**PIC imaging** The setup is derived from(27) (Fig. 1A). The samples were mounted onto an inverted microscope equipped with a 100x oil immersion objective (NA=1.4) and a 3D piezo-scanner. An horizontally polarized beam of He-Ne laser (633nm wavelength) is split into two beams (herein called probe and reference beams) by a Wollaston prism and sent to the microscope objective through a telecentric lens system. The transmitted probe and reference beams are collected by a microscope objective (NA=0.8) and imaged on a mirror (at normal incidence) by a 100mm focal length lens, forming an afocal system. The retroreflected beams having the same optical path as the incident ones, recombine in a Wollaston prism. The vertically polarized recombination reflected by the polarizing cube is sent to a fast photodiode. Heating of the nanoparticles is performed by a 514 nm wavelength beam from an Argon laser which intensity is modulated at high frequency (900kHz) by an acousto-optical modulator. We performed resolutions studies in two different heating beam configurations: a focused illumination where the green beam is superimposed with the probe beam and an epi-illumination where the green beam is defocused at the "entrance" of the objective and a wide region of the sample is illuminated.

The heating beam induces a periodic phase difference between the probe and reference beams when the red probe beam is positioned on a heated nanoparticle. This gives rise to a modulation of the red intensity which is filtered from green stray light of the heating beam and detected with a fast photodiode. The photodiode signal is sent to a lock-in amplifier to detect the phase difference between the probe and reference beams at the same frequency as the heating beam intensity modulation. A lock-in amplifier is used to detect the phase difference at the modulation frequency with an integration time of 10 ms. Microscopic images are obtained by scanning the sample with a XYZ piezo-electric translation stage.

**Cell culture, transfection of COS7 cells and immunostaining.** COS7 cells were cultured plated on #1 glass slides in DMEM medium supplemented with streptomycin (100 µg/ml), penicillin (100 U/ml), and 10 % bovine serum in a humidified atmosphere (95 %) at 5 % $CO_2$ and $37^o$. Cells were used for 12-14 passages and were transferred every 4 days. Transfection was performed using FUGEN. Cells exhibiting confluence of ~ 30% were used for transfection with 10 µg for 1 ml volume of cDNA coding for a metabotropic receptor for glutamate containing of myc-tag at the extracellular N-terminus (mGluR5-myc(28)). Transfection efficiency was on the order of 50%. After 12h, the cells were fixed using the following protocol: 10min in 4% paraformaldehyde / sucrose, 15min in phosphate saline buffer (PBS) with 50nM NH4Cl, then 3 rinses in PBS with 0.3% Bovine Serum Albumine (BSA). Then, a first immunostaining was performed using antimyc antibodies tagged with alexa568 dyes (herein named amyc-Alexa568, 20min at room temperature 10µg/ml, 0.3% BSA). After two rinses in PBS, a secondary immunostaining by antiIgG-10nm gold (goat anti-mouse to label the amyc-Alexa568, Auroprobes Amersham, 20min at room temperature, 0.3 % BSA) was performed for different concentrations (10µg/ml to 0.1µg/ml) followed by 3 rinses in PBS.

**Fluorescence and scattering imaging.**

Fluorescence images are recorded using the focused green heating beam for excitation. The power was reduced to ~ 10µW. The fluorescence is collected with the high NA objective, filtered with a high pass filter (OG550) and imaged in a confocal configuration onto an avalanche photodiode. Scattering images are recorded with the modulated heating beam in absence of the red beams. The scattered green light is detected by the fast photodiode (red-pass filter removed) and demodulated with the lock-in amplifier.

**RESULTS AND DISCUSSION**

**Signal calculation, resolution**

In this section, we derive an analytical form of the modulated signal detected by the photodiode. We chose a coordinate system with an origin at the focal point of the heating beam which coincides with the one of the two red beams, the probe beam (Fig 1B). Since we use small absorbing nanoparticles, we can make the assumption of a punctual heat source with a modulated heat power $\sigma I(\vec{r}_0) \cdot [1 + \cos(\omega t)]$ where $\sigma$ is the absorption cross section of the nanoparticle and $I(\vec{r}_0)$ the intensity of the green beam at its position (figure 1.b). The temperature rise $\Delta T$ at a position $\vec{r} = (x, y, z)$ induced by this punctual heat source in a homogeneous medium is derived from the equation of heat conduction (29) and is given by

$$\Delta T(|\vec{r} - \vec{r}_0|, \vec{r}_0, t) = \frac{\sigma \times I_{heat}(\vec{r}_0)}{4\pi\kappa |\vec{r} - \vec{r}_0|} \left[ 1 + \cos\left(\omega t - \frac{|\vec{r} - \vec{r}_0|}{R}\right) e^{-\frac{|\vec{r} - \vec{r}_0|}{R}} \right].$$

$R = \sqrt{2\kappa/\omega C}$ is the characteristic length for heat diffusion at ω, where κ is the thermal conductivity of the medium (0.19 W.K$^{-1}$.m$^{-1}$ for water) and C its heat capacity for a unit volume. At the working modulation frequency, R is on the order of the probe beam transverse focal spot size (27). The temperature rise induces a change in the refractive index medium at position $\vec{r}$:

$$\Delta n(|\vec{r} - \vec{r}_0|, \vec{r}_0, t) = \frac{\partial n}{\partial T} \cdot \Delta T(|\vec{r} - \vec{r}_0|, \vec{r}_0, t)$$
$$= I_{heat}(\vec{r}_0) \cdot f(|\vec{r} - \vec{r}_0|, t).$$

$\frac{\partial n}{\partial T}$ is the variation of the refractive index per unit of temperature (~10$^{-4}$ K$^{-1}$ for water at room temperature(30)). The heat diffusion characteristic length R being on the order of the optical wavelength, diffraction effects play important roles. The exact expression of the emerging probe beam can in principle be derived using light scattering theory(31). The solution is complicated, and is out of the scope of this paper.

To have a simple qualitative physical insight we shall consider now the case were the geometrical optics is valid. For weak refractive index changes, we use the thin-phase grating approximation by assuming that the phase of the optical field can be calculated along the optical rays as they would be in the absence of phase-shifting object. The global phase-shift is thus obtained by accumulating along unperturbed optical rays the variation of refractive index through the entire object. When the object is located in regions where the probe wave front is plan, the signal of the interference between the probe and the beams, detected at the modulation frequency ω, writes:

$$S(x_0, y_0, z_0) = \eta \iint_\infty dxdy\, 2I_{probe}(x, y, z_0) \cdot \int_l dz\, \frac{2\pi}{\lambda} \Delta n_\omega\left(\left|\vec{r} - \vec{r}_0\right|, \vec{r}_0\right)$$

$$= \frac{4\pi}{\lambda} \eta I_{heat}(x_0, y_0, z_0) \iint_\infty dxdy\, I_{probe}(x, y, z_0) \cdot \int_l dz\, f_\omega\left(\left|\vec{r} - \vec{r}_0\right|\right)$$

where η is the conversion factor of the detection chain and *l* is a distance on which the variations of refractive index are significant *(l>R)*. The index $\omega$ in $f_\omega$ refers to the component of *f* at frequency $\omega$ detected by the lock-in amplifier. We assume that the probe and reference beams have the same intensity profiles so that only $I_{probe}$ appears in the equations.

This latter formulae gives the origin of the resolution of the PIC method when we scan the sample i.e. the particle position $\vec{r}_0$. In the transverse plane (xOy), it is given by the product of the heating profile with the convolution between the probe beam and the phase shifting object profiles. However, the axial resolution in the plane probe wave-front approximation is given by the product of the heating profile by the probe beam profile.

In order to estimate the resolution of the PIC method and the probe-reference optical phase shift, the modulated refractive index variations $\Delta n_\omega(\vec{r}, \vec{r}_0)$ are modeled by a "phase-sphere" of radius $\rho_{Th}$ centred on the heated particle with a uniform modulated refractive index ($\overline{\Delta n_\omega}(\vec{r}_0)$). We define $\rho_{Th}$ and $\overline{\Delta n_\omega}(\vec{r}_0)$ as:

$$\rho_{Th} = \frac{\int_0^\infty d^3\vec{r}\, r \cdot \Delta n_\omega(\vec{r}, \vec{r}_0)}{\int_0^\infty d^3\vec{r}\, \Delta n_\omega(\vec{r}, \vec{r}_0)} = \sqrt{2}R, \qquad \overline{\Delta n}(\vec{r}_0) = \frac{\int_0^{\rho_{Th}} d^3\vec{r}\, \Delta n_\omega(\vec{r}, \vec{r}_0)}{\int_0^{\rho_{Th}} d^3\vec{r}} \approx 0.34 \frac{\partial n}{\partial T} \frac{\sigma}{4\pi\kappa R} I(\vec{r}_0)$$

In this case, the detected signal writes:

$$S(x_0, y_0, z_0) = \frac{4\pi}{\lambda} \eta \overline{\Delta n}(\vec{r}_0) \iint_\infty dx\,dy\, I_{probe}(x, y, z_0) \sqrt{\rho_{Th}^2 - (x-x_0)^2 - (y-y_0)^2} \quad (1)$$

The maximum phase shift $\Delta\phi_{max}$ is obtained when the particle is at the focal point of the heating beam ($\vec{r}_0 = \vec{0}$). In the case where the probe beam profile is nearly uniform over the "phase-sphere", this phase shift is given by:

$$\Delta\phi_{max} = \frac{2\pi}{\lambda} \overline{\Delta n_\omega}(\vec{0}) \frac{2.\int_0^{\rho_{Th}} \sqrt{\rho_{Th}^2 - \rho^2}\, 2\pi\rho\, d\rho}{\int_0^{\rho_{Th}} 2\pi\rho\, d\rho} = \frac{16\pi}{3} \overline{\Delta n_\omega}(\vec{0}) \frac{\rho_{Th}}{\lambda}$$

For a 10 nm particle absorbing an average heating power of 300nW modulated at a frequency of 1 MHz, we find $\Delta\phi_{max} \approx 16\cdot 10^{-5}\, rad$.

**Resolution and Phase-shift measurements**

The transverse and axial resolutions of the PIC method were studied on samples of 10nm gold colloidal nanoparticles embedded in a PVA film (few tens of nm thick) spin-coated on glass slide. We used a heating beam which fulfilled the back aperture of the microscope objective such that its transverse dimension was $1.22\, \lambda_{heat}/2\mathrm{NA} = 224\, nm$ (FWHM) and its axial dimension $1.22\, \pi\, \lambda_{heat}/\mathrm{NA}^2 = 1.0\, \mu m$ (FWHM). However the probe and reference beams under-filled the aperture of the objective. Its transverse size at the focal plane was 550±50 nm which leads to a longitudinal size 5.0±0.9µm. As mentioned above, the theoretical transverse resolution of the PIC method involves the profiles of the heating and probe beams.

For simplification, assuming Gaussian beams at the objective focal plane and a uniform phase profile in a sphere of $2\rho_{Th} = 370 nm$ diameter, equation (1) leads to a theoretical transverse (resp. axial) resolution of 209±2 nm (resp. 0.85±0.05 µm) in the well focussed heating configuration and 530±70nm (resp. 5.0±0.9µm) in the uniform heating one.

In the focused heating beam configuration, we measured a transverse resolution of 215±5 nm (Fig. 2A&B) and an axial resolution of 1.2±0.2 µm (Fig. 2A&C). And as expected, the resolution degraded in the uniform heating configuration and is 460±25 nm (Fig 2D) in the transverse plane and is 6.0±0.5 µm for the axial direction (Fig 2E).

In both configurations the measurements are in fairly good agreement with what expected from our simple theoretical analysis.

The pointing accuracy at which single nanogold particles can be localized with the PIC method is potentially very high. Since photothermal signals of metal particles are stable and do not saturate, single nanoparticles can be imaged with very high signal-to-noise ratio (SNR). In a typical situation, 10 nm gold particles absorbing 300nW heating power are detected with a peak SNR greater than 30 when the integration time is 10 ms per pixel. Taking the image pixel size equal to the standard deviation of the PIC resolution(32), the expected pointing accuracy is then 10 nm in the transverse directions, far below the optical diffraction limit(32-34). This pointing accuracy can be dramatically improved with higher SNRs(32-34), using for example larger heating intensities or longer integration times. This very high precision localization of nanoparticles could be of considerable interest for the precise localisation of proteins in cells.

In order to measure the phase-shift $\Delta\phi$ induced by heating a single particle we insert an electro-optic modulator (EOM) between the two lenses of the telecentric system. The axis of the EOM are set parallel to the respective polarisations of the reference and the probe beams. Applying a modulated voltage to the EOM induces a phase-shift which mimics the phase-shift produced by a heated particle. $\Delta\phi$ is thus deduced from the amplitude of the modulated voltage obtained when the demodulated signals equal that of the heated single particles.

For 10 nm gold spheres absorbing an average heating power of 300nW modulated at a frequency of 1 MHz, we measured $\Delta\phi \approx 8 \pm 2\,10^{-5} rad$ in qualitative agreement with our theoretical estimation.

**Immunohistochemistry assay**

We performed an immunohistochemistry assay to demonstrate the capabilities of the PIC method to image low amounts of immunogold labeled proteins in cells. Protein conjugated colloid gold particles can be obtained commercially in sizes ranging from one nanometer to hundreds of nanometers. Electron microscopy prefers to use metal particles smaller than 10 nm because of their better penetration in cellular organelles(35). We used COS7 cells transfected with membrane protein mGlurR5-myc (see methods). The membrane proteins were primarily labeled with Anti-myc antibodies tagged with alexa568 dyes. As a secondary labeling stage, we used antiIgG-10nm gold antibodies (see methods). Fluorescence images easily discriminate cells that contained fluorescence-labeled receptors from untransfected ones (Fig. 3B, E&H). Specificity of the gold labeling was subsequently ensured as no signal was "usually" detected by the PIC method on untransfected cells as shown on Fig. 3A-C (less than $4.10^{-3}$ particles per focal spot area ~$0.07\mu m^2$ for untransfected cell (N=10 cells), compared to much more than one particle per $0.07\mu m^2$ in transfected cell (N=28 cells) for the same labeling conditions). Photothermal images of transfected cells are shown on Fig. 3 for two concentrations of antiIgG-10nm gold (10µg/ml in Fig. 3D-F and 0.5µg/ml in Fig. 3G-I). AntiIgG-10nm gold labeling is clearly observed on the cell membrane. The images are identical when the same regions are recorded successively due to the non-photbleaching feature of the labels. Three dimensional pictures of the membrane protein labeling can thus be obtained by changing the axial position of the samples (supplementary materials). Resolved discrete spots which may be attributed to single nanoparticles can be observed in cell regions with low density of labels as in the inset of Fig. 3I, (recorded with a high scan resolution, 40nm). We further decreased the labeling concentration of antiIgG-10nm gold to 0.1µg/ml in order to obtain a sparse labeling on the cell membranes (Fig. 4B-C). Several arguments indicate that we achieved single 10nm gold particle detection in cells. First, when increasing the intensity of the heating laser by 5 folds, no weaker spot appears in the image (see Figure 4B-C). This indicates that we detect the smallest absorbing objects present on the plasma membrane of the cells. Second, we constructed and analysed the histograms of the signal heights measured on three different samples: bare 10 nm gold colloids in a spin-coated PVA film (Fig4D), antiIgG-10nm gold antibodies also embedded in a PVA (Fig4E) and membrane receptors labeled with a dilute solution (0.1µg/ml) of antiIgG-10nm gold (Fig4F).

The fairly narrow unimodal distribution obtained in Fig4D confirms that the spots stem from individual nanospheres(27). The width of the distribution is equal to 0.3 (center normalized to 1). It arises from the dispersion of the particle sizes (10% given by the manufacturer) which translates to 30% dispersion in the signal heights since absorption cross-section scales as the volume of the nanosphere. Interestingly, the distribution obtained from spin coated antiIgG-10nm gold antibodies is bimodal. It is well fitted by two Gaussian curves centred at signal heights 1 (normalisation value) and 2.0±0.1, with respective widths $\delta=0.32\pm0.03$ and $0.4\pm0.1 \approx \sqrt{2}\times\delta$. The relative positions and widths of the two peaks indicate that the spots arise from either one (63±10%) or two metal particles. Assuming a Poisson distribution, the mean number of nanoparticles per antibody is thus equal to one. However, we would expect to detect 13% of antibodies labeled with more than 2 nanoparticles which is not the case as stronger signals that could have been attributed to more than two particles were extremely rare. Steric limitation can account for this as the typical size of a IgG is of the order of the size of the nanoparticles.

The histogram constructed from PIC images of cells (Fig4F) is well fitted by four Gaussian curves. The widths of the Gaussian curves were set to SQRT(N)×δ, N being the number of the peak and δ a free fit parameter (36). The average signal height of the peaks scales linearly with the peak number (inset on Fig4G). By comparison with the results obtained from spin-coated antiIgG-10nm gold antibodies one can conclude that single antiIgG-10nm gold antibodies are detected. The first peak corresponds to single IgG antibodies labeled with only one nanoparticle. Altogether the data presented on Fig4 indicates that individual 10 nm nanoparticles can be detected on the plasma membrane of cells. The SNR at which the individual 10 nm nanoparticles are detected depends on the heating beam intensity. For this study (Fig. 3 and 4), the intensity was of the order of $3MW/cm^2$ which corresponds to a power of ~150nW absorbed by a nanoparticle (in an aqueous environment). The SNR was then better than 20

**CONCLUSION**

We demonstrated in this work that 10nm particles commonly used to label proteins in cells can be imaged using an all optical method. We measured the transverse and axial resolution of the PIC method and compared the measurements to a simple model.

The advantages of the PIC method over fluorescent methods for single molecule detection arise from the absence of saturation of the signals, from the absence of the autofluorescence or scattering due to the environment -or to the cells themselves- and from the absence of blinking and photobleaching of the labels. The latter limitation is very restrictive to perform 3D localization of the single fluorescent molecules as multiple records of the same molecule are hardly feasible. On the contrary, the stability of the PIC signals provides a way to localize in 3D single particles with a number of recordings and signal levels which can be arbitrary high. Single nanoparticles can thus be localized in the scattering environment formed by a cell with very high pointing accuracy. Our work show that it is in principle possible to multiplex fluorescence and PIC images. Spots from a PIC image can be accurately located with respect to specific cellular organelles fluorescently labeled. Furthermore, nanoparticles with a shifted plasmon resonance (of different shape or composition) could be used to perform multiple color PIC imaging.

For live biological samples, the SNRs might however be limited by the admissible temperature rise in the sample. For particles detected on the cell membranes with a SNR of 10, we estimated a temperature increase of 15K on the surface of the 10 nm gold particles (27). This temperature rise decreases as the reciprocal distance from the particle center.

The high sensitivity of the method allowed precise quantification of the extremely low number (1 or 2) of 10 nm gold colloids present on commercially available IgG antibodies. This could be of great interest when low expression proteins are aimed to be detected. The method does not require amplification to achieve high sensitivity, giving an unbiased picture of the expression pattern of the cell. More generally, the PIC method is accurate and robust enough to perform stoechiometry of gold nanoparticles as commonly used in protein and DNA chips.

We thank Dr. Dionysia Theodosis for the gift of the antiIgG-10nm gold antibodies and Pr. Michel Orrit for stimulating discussions. This work was supported by grants from the CNRS, the Conseil Régional d'Aquitaine and the ministère de la recherche.

The authors declare that they have no competing interests.

1.  Special Issue (1999) *Science* **283**.

2.  Deschenes, L. A. & Vanden Bout, D. A. (2001) *Science* **292,** 255-258.

3.  Sako, Y., Minoghchi, S. & Yanagida, T. (2000) *Nat Cell Biol* **2,** 168-72.

4.  Schutz, G. J., Kada, G., Pastushenko, V. P. & Schindler, H. (2000) *EMBO J* **19,** 892-901.

5.  Harms, G. S., Cognet, L., Lommerse, P. H., Blab, G. A., Kahr, H., Gamsjager, R., Spaink, H. P., Soldatov, N. M., Romanin, C. & Schmidt, T. (2001) *Biophys J* **81,** 2639-46.

6.  Iino, R., Koyama, I. & Kusumi, A. (2001) *Biophys. J.* **80,** 2667-2677.

7.  Seisenberger, G., Ried, M. U., Endress, T., Buning, H., Hallek, M. & Brauchle, C. (2001) *Science* **294,** 1929-1932.

8.  Vanden Bout, D. A., Yip, W.-K., Hu, D., Fu, D., Swager, T. M. & Barbara, P. (1997) *Science* **277,** 1074-1077.

9.  Lu, H. P., Xun, L. & Xie, X. S. (1998) *Science* **282,** 1877-82.

10. Schuetz, G. J., Schindler, H. & Schmidt, T. (1997) *Biophys.J.* **73(2),** 1073-1080.

11. Ha, T., Enderle, T., Chemla, D. S., Selvin, P. R. & Weiss, S. (1996) *Phys.Rev.Lett.* **77,** 3979-3982.

12. Lacoste, T. D., Michalet, X., Pinaud, F., Chemla, D. S., Alivisatos, A. P. & Weiss, S. (2000) *Proc Natl Acad Sci U S A* **97,** 9461-9466.

13. Zhuang, X., Bartley, L. E., Babcock, H. P., Russell, R., Ha, T., Herschlag, D. & Chu, S. (2000) *Science* **288,** 2048-2051.

14. Dubertret, B., Skourides, P., Norris, D., Noireaux, V., Brivanlou, A. & Libchaber, A. (2002) *Science* **298,** 1759-62.

15. Wu, X., Liu, H., Liu, J., Haley, K., Treadway, J., Larson, J., Ge, N., Peale, F. & Bruchez, M. (2003) *Nat Biotechnol* **21,** 41-46.

16. Jaiswal, J., Mattoussi, H., Mauro, J. & Simon, S. (2003) *Nat Biotechnol* **21,** 47-51.

17. Larson, D. R., Zipfel, W. R., Williams, R. M., Clark, S. W., Bruchez, M. P., Wise, F. W. & Webb, W. W. (2003) *Science* **300,** 1434-1436.

18. Nirmal, M., Dabbousi, B. O., Bawendi, M. G., Macklin, J. J., Trautman, J. K., Harris, T. D. & Brus, L. E. (1996) *Nature* **383,** 802-804.

19. Sheetz, M. P., Turney, S., Qian, H. & Elson, E. L. (1989) *Nature* **340,** 284-8.
20. Baschong, W., Lucocq, J. M. & Roth, J. (1985) *Histochemistry* **83,** 409-11.
21. Slot, J. W. & Geuze, H. J. (1985) *Eur J Cell Biol* **38,** 87-93.
22. Frey, P. A. & Frey, T. G. (1999) *J Struct Biol* **127,** 94-100.
23. Hainfeld, J. F. & Powell, R. D. (2000) *J Histochem Cytochem* **48,** 471-80.
24. Schultz, S., Smith, D. R., Mock, J. J. & Schultz, D. A. (2000) *Proc Natl Acad Sci U S A* **97,** 996-1001.
25. Gelles, J., Schnapp, B. J. & Sheetz, M. P. (1988) *Nature* **331,** 450-3.
26. Sönnichsen, C., Geier, S., Hecker, N. E., von Plessen, G., Feldmann, J., Ditlbacher, H., Lamprecht, B., Krenn, J. R., Aussenegg, F. R., Chan, V. Z.-H., Spatz, J. P. & Möller, M. (2000) *Applied Physics Lettres* **77,** 2949-2951.
27. Boyer, D., Tamarat, P., Maali, A., Lounis, B. & Orrit, M. (2002) *Science* **297,** 1160-1163.
28. Serge, A., Fourgeaud, L., Hemar, A. & Choquet, D. (2002) *J. Neurosci.* **22,** 3910-3920.
29. Carslaw, H. S. & Jaeger, J. C. (1993) *Conduction of heat in solids*, Oxford).
30. Tilton, L. W. & Taylor, J. K. (1938) *J. Res. Natl. Bur. Stand.* **20**.
31. Chu, B. (1974) . *Laser Light Scattering* (Academic Press, N.Y.).
32. Thompson, R. E., Larson, D. R. & Webb, W. W. (2002) *Biophys. J.* **82,** 2775-2783.
33. Bobroff, N. (1986) *Rev. Sci. Instrum* **57,** 1152–1157.
34. Schmidt, T., Schuetz, G. J., Baumgartner, W., Gruber, H. J. & Schindler, H. (1996) *Proc.Natl.Acad.Sci.U.S.A.* **93(7),** 2926-2929.
35. Hayat, M. A. (1989) *Colloidal Gold: Principles, Methods and Applications* (Academic, San Diego).
36. Schmidt, T., Schutz, G. J., Gruber, H. J. & Schindler, H. (1996) *Analytical Chemistry* **68,** 4397-4401.

Figure captions:

Figure 1:

(A) Schematic diagram of the optical setup. The heating beam (in blue) heats the metallic nanoparticle. The modulated red beam is split into two probe beam by a wollaston prism which are retroreflected by an afocal system to be recombined by the wollaston prism. After demodulation of one output of this interferometer, the modulated phase shift induced by the heating present on one arm of the interferometer can be detected when a nanoparticle is present. Fluorescent labels excited by the heating beam can also be detected on the same setup using the Single Photon Counting avalanche photodiode. (B) Coordinate system with an origin at the focal point of the heating beam which coincides with the one of the two red beams, the probe beam.

Figure 2:

Resolution of the PIC method. (A) 3D representation of the signal obtained from a single 10nm gold nanosphere in PVA. (B) Transverse resolution with a well focussed heating beam. Fitting the signal by a Gaussian curve gives an axial resolution 215±5 nm. (C) same as (B) but with a uniform heating beam. The transverse resolution is degraded: 460±25 nm (FWHM). (D) Axial resolution in the well focussed situation: 1.2±0.2 µm (FWHM) (E) same as (B) but with a uniform heating beam. The resolution is 6.0±0.5 µm (FWHM).

Figure 3:

Scattering (A, D & G), fluorescence (B, E & H) and photothermal (C, F & I) images recorded on COS7 cells. (A-C) correspond to untransfected cells whereas (D-I) correspond to cells expressing a membrane protein (mGluR5, a receptor for neurotransmitter) containing a myc tag. All cells were immuno-labeled with amyc-Alexa568 (10µg/ml) and with antiIgG-10nm gold as secondary antibody, for two different concentrations (10µg/ml in Fig. 3A-F and 0.5µg/ml in Fig. 3G-I). Cells expressing mGluR5 were efficiently labeled with antimyc-Alexa568 and antiIgG-10nm gold. Inset of (I) shows a detail of the PIC image revealing individual antiIgG-10nm gold imaging. The heating beam intensity is 3MW/cm$^2$.

Figure 4:

Evidences for single 10nm nanoparticles detection in the cells. (A) Fluorescence image recorded on a portion of a transfected cell immuno-labeled with amyc-Alexa568 (10µg/ml) and antiIgG-10nm gold as secondary antibody (0.1 µg/ml). (B-C): A sparse labeling is obtained in the PIC images. The heating intensity is increased from 3 MW/cm² (B) to 15 MW/cm² (C), but no new spots are observed in C (the two images are displayed with the same intensity scale). (D-F): Histogram of the signals for different samples: (D) 10 nm gold colloids spin coated in PvOH films. The distribution is well fitted by a single Gaussian curve as expected for single nanoparticles. (E) antiIgG-10nm gold antibodies spin coated in PvOH films. The distribution is well fitted by a two Gaussian curves. The position of the first maxima being set to 1, that of the second maxima is 2.0±0.1 and its width ~√2 greater than that of the first one. This indicates that single particles are detected in the first peak and two particles in the second. IgG antibodies are thus labeled with either one or two 10 nm gold colloids, but not more. (F) antiIgG-10nm gold antibodies detected on the cell membrane of transfected cells as in Fig. 4B. The distribution is fitted by 4 Gaussian curves. The positions of the Gaussian fits maxima follow a linear law (G) indicating that particles in the first peak correspond to single nanoparticles.

Figure 1

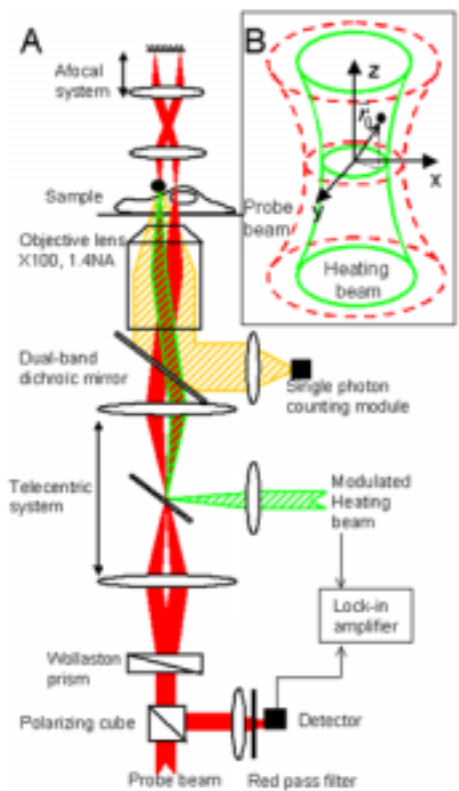

Figure 2

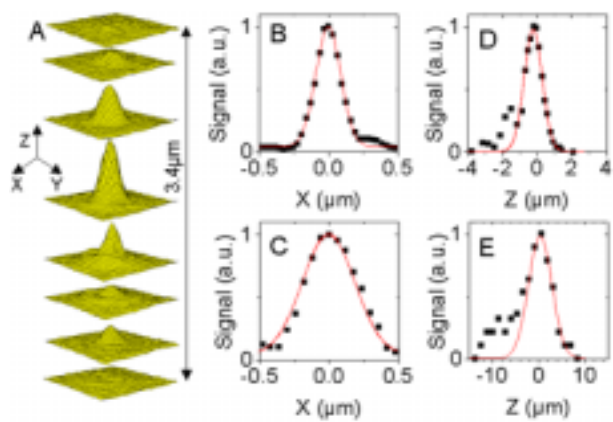

Figure 3

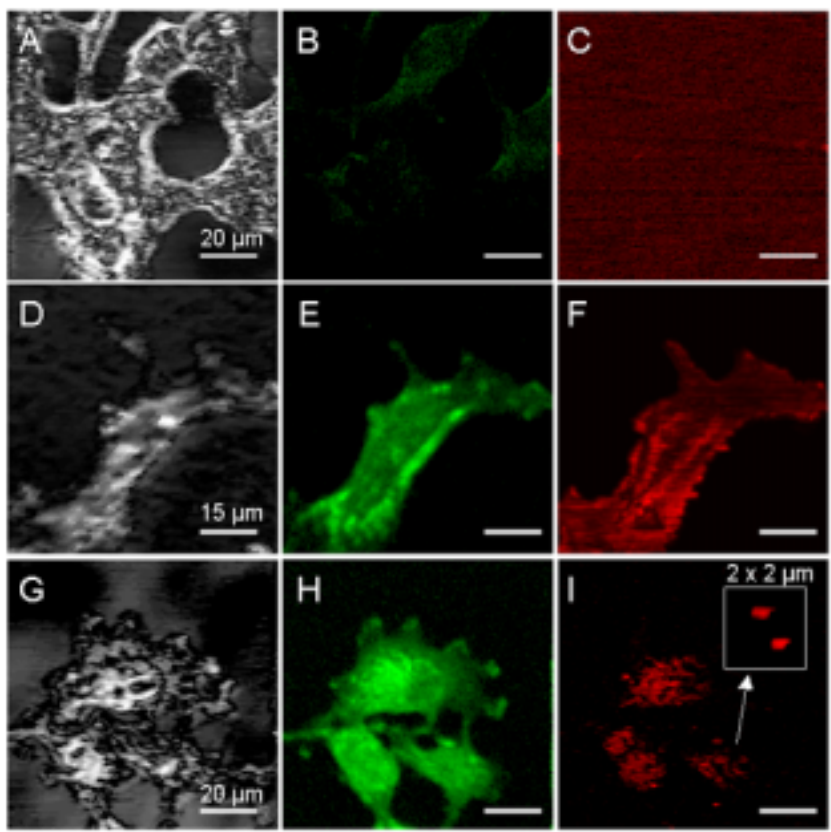

Figure 4

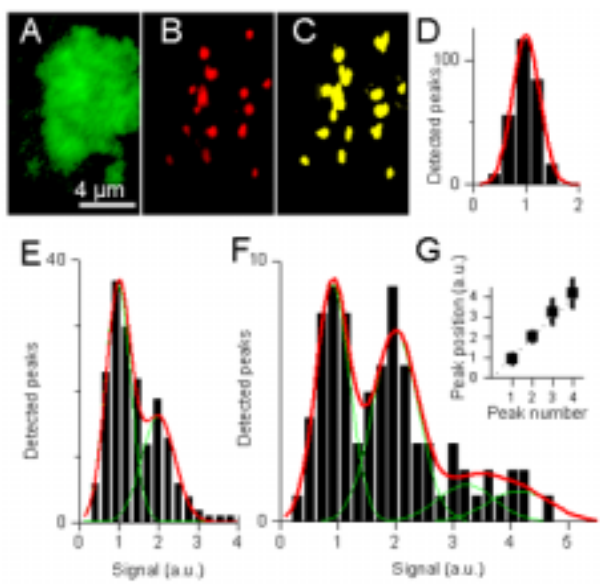